\font\bbb=msbm10                                                   

\def\C{\hbox{\bbb C}}

\def\R{\hbox{\bbb R}}

\def\IJMPC{{\sl Int.\ J. Mod.\ Phys.\ C}}

\def\JPA{{\sl J. Phys.\ A}}

\def\JSP{{\sl J. Stat.\ Phys.}}
\def\LAS{{\sl Los Alamos Science}}
\def\MPCPS{{\sl Math.\ Proc.\ Camb.\ Phil.\ Soc.}}

\def\PD{{\sl Physica D}}

\def\PR{{\sl Phys.\ Rev.}}
\def\PRA{{\sl Phys.\ Rev.\ A}}

\def\PRE{{\sl Phys.\ Rev.\ E}}
\def\PRL{{\sl Phys.\ Rev.\ Lett.}}
\def\PRSLA{{\sl Proc.\ Roy.\ Soc.\ Lond.\ A}}

\def\Sc{{\sl Science}}

\def\dajm{\hbox{D. A. Meyer}}

\def\brosl{\hbox{B. Hasslacher}}

\def\bogtay{\hbox{B. M. Boghosian and W. Taylor, IV}}

\def\hfb{\hfil\break}

\catcode`@=11
\newskip\ttglue

   \font\ninerm=cmr9    \font\eightrm=cmr8   \font\sixrm=cmr6
  \font\ninebf=cmbx9   \font\eightbf=cmbx8  \font\sixbf=cmbx6
  \font\nineit=cmti9   \font\eightit=cmti8  
  \font\ninesl=cmsl9   \font\eightsl=cmsl8  
  \font\ninemi=cmmi9   \font\eightmi=cmmi8  \font\sixmi=cmmi6

\font\bigtenbf=cmr10 scaled\magstep2 

\def\ninepoint{\def\rm{\fam0\ninerm}%
  \textfont0=\ninerm \scriptfont0=\sixrm
  \textfont1=\ninemi \scriptfont1=\sixmi
  \textfont\itfam=\nineit  \def\it{\fam\itfam\nineit}%
  \textfont\slfam=\ninesl  \def\sl{\fam\slfam\ninesl}%
  \textfont\bffam=\ninebf  \scriptfont\bffam=\sixbf
    \def\bf{\fam\bffam\ninebf}%
  \tt \ttglue=.5em plus.25em minus.15em
  \normalbaselineskip=11pt
  \setbox\strutbox=\hbox{\vrule height8pt depth3pt width0pt}%
  \normalbaselines\rm}

\def\eightpoint{\def\rm{\fam0\eightrm}%
  \textfont0=\eightrm \scriptfont0=\sixrm
  \textfont1=\eightmi \scriptfont1=\sixmi
  \textfont\itfam=\eightit  \def\it{\fam\itfam\eightit}%
  \textfont\slfam=\eightsl  \def\sl{\fam\slfam\eightsl}%
  \textfont\bffam=\eightbf  \scriptfont\bffam=\sixbf
    \def\bf{\fam\bffam\eightbf}%
  \tt \ttglue=.5em plus.25em minus.15em
  \normalbaselineskip=9pt
  \setbox\strutbox=\hbox{\vrule height7pt depth2pt width0pt}%
  \normalbaselines\rm}

\def\sfootnote#1{\edef\@sf{\spacefactor\the\spacefactor}#1\@sf
      \insert\footins\bgroup\eightpoint
      \interlinepenalty100 \let\par=\endgraf
        \leftskip=0pt \rightskip=0pt
        \splittopskip=10pt plus 1pt minus 1pt \floatingpenalty=20000
        \parskip=0pt\smallskip\item{#1}\bgroup\strut\aftergroup\@foot\let\next}
\skip\footins=12pt plus 2pt minus 2pt
\dimen\footins=30pc

\def\ie{{\it i.e.}}
\def\eg{{\it e.g.}}

\def\@versim#1#2{\lower.5pt\vbox{\baselineskip0pt \lineskip-.5pt
    \ialign{$\m@th#1\hfil##\hfil$\crcr#2\crcr\sim\crcr}}}
\def\gsim{\mathrel{\mathpalette\@versim\>}}

\magnification=1200
\input epsf.tex

\dimen0=\hsize \divide\dimen0 by 13 \dimendef\chasm=0
\dimen1=\hsize \advance\dimen1 by -\chasm \dimendef\usewidth=1
\dimen2=\usewidth \divide\dimen2 by 2 \dimendef\halfwidth=2
\dimen3=\usewidth \divide\dimen3 by 3 \dimendef\thirdwidth=3
\dimen4=\hsize \advance\dimen4 by -\halfwidth \dimendef\secondstart=4
\dimen5=\halfwidth \advance\dimen5 by -10pt \dimendef\indenthalfwidth=5
\dimen6=\thirdwidth \multiply\dimen6 by 2 \dimendef\twothirdswidth=6
\dimen7=\twothirdswidth \divide\dimen7 by 4 \dimendef\qttw=7
\dimen8=\qttw \divide\dimen8 by 4 \dimendef\qqttw=8
\dimen9=\qqttw \divide\dimen9 by 4 \dimendef\qqqttw=9

\parskip=0pt\parindent=0pt

\line{to appear in \JPA\hfil 15 November 2000}
\line{\hfil quant-ph/0105087}
\bigskip\bigskip\bigskip
\centerline{\bf\bigtenbf FROM GAUGE TRANSFORMATIONS}
\bigskip
\centerline{\bf\bigtenbf TO TOPOLOGY COMPUTATION} 
\bigskip
\centerline{\bf\bigtenbf IN QUANTUM LATTICE GAS AUTOMATA}
\vfill
\centerline{\bf David A. Meyer}
\bigskip 
\centerline{\sl Project in Geometry and Physics}
\centerline{\sl Department of Mathematics}
\centerline{\sl University of California/San Diego}
\centerline{\sl La Jolla, CA 92093-0112}
\centerline{\tt dmeyer@chonji.ucsd.edu}
\smallskip
\centerline{and}
\smallskip
\centerline{\sl Institute for Physical Sciences}
\centerline{\sl Los Alamos, NM 87544}
\vfill
\centerline{ABSTRACT}
\bigskip
\noindent The evolution of a quantum lattice gas automaton (LGA) for a 
single charged particle is invariant under multiplication of the wave 
function by a global phase.  Requiring invariance under the 
corresponding local gauge transformations determines the rule for 
minimal coupling to an arbitrary external electromagnetic field.  We 
develop the Aharonov-Bohm effect in the resulting model into a 
{\sl constant\/} time algorithm to distinguish a one dimensional 
periodic lattice from one with boundaries; any classical deterministic 
LGA algorithm distinguishing these two spatial topologies would have 
expected running time on the order of the cardinality of the lattice.

\bigskip
\global\setbox1=\hbox{PACS numbers:\enspace}
\global\setbox2=\hbox{PACS numbers:}
\parindent=\wd1
\item{PACS numbers:}  03.65.Lx,  
                      03.65.Vf,  
                      03.65.Pm.  

\item{\hbox to \wd2{KEY\hfill WORDS:}}
                      quantum lattice gas; quantum computation;
                      Aharonov-Bohm effect; 
\item{}               spatial topology.

\vfill
\eject

\headline{\ninepoint\it Topology computation     \hfil David A. Meyer}
\parskip=10pt
\parindent=20pt

Quantum lattice gas automata (QLGA) have been proposed as a possible
architecture for solid state quantum computers since they require only
an array of sites which can support an extended (multi-)electron wave 
function [1].  The simplicity of such an architecture makes nanoscale 
fabrication plausible [2].  The main incentives for pursuing the 
program of quantum computation, however, are the quantum algorithms of
Shor [3] and Grover [4], for example, which provide substantial
improvements over deterministic or probabilistic algorithms.  These
quantum algorithms can be efficiently implemented with quantum gate 
arrays [5], but at least in simple translations---analogous to 
deterministic billiard ball models for universal computation 
[6]---seem impractical to implement in QLGA models.  Just as 
deterministic LGA efficiently simulate fluid flow in certain parameter 
regimes [7], QLGA seem best suited for simulation of quantum physical 
processes [1,8].  Single particle QLGA have been shown,
in fact, to limit to the Schr\"odinger [9] and Dirac [1] equations
under appropriate conditions.  In this letter we show that in 
simulating a single quantum particle, QLGA efficiently perform a new 
and interesting computation---of spatial topology.

We draw inspiration from the topological invariance of the spectral 
flow in a one parameter family of Dirac operators [10].  Spectral 
flow derives from the components of the group of gauge 
transformations, so after recalling the definition of one dimensional 
single particle QLGA with inhomogeneous potentials, we analyze which 
evolution rules are gauge equivalent.  We find a difference between 
the gauge equivalence classes of rules on periodic lattices and 
lattices with boundary, and show how this may be exploited to 
distinguish between these two spatial topologies.  We conclude with a 
discussion of the complexity of this quantum computation and point out 
directions for further investigation.

A one particle QLGA is a discrete time model for a quantum particle
moving in an array (lattice) of sites.  Here we will consider only
finite one dimensional lattices $L$, with or without periodic boundary
conditions.  In these cases we need only allow two velocities 
$\{\pm1\}$ in order to construct a model which limits to the 
Schr\"odinger [9] or Dirac [1] equation as the timestep and lattice
spacing scale to zero appropriately.  The amplitudes for the particle
to be (left,right) moving at a lattice site $x \in L$ combine into a
two component complex vector 
$\psi(t,x) := \bigl(\psi_{-1}(t,x),\psi_{+1}(t,x)\bigr)$ which evolves
as
$$
\psi(t+1,x) = w_{-1}(x) \psi(t,x-1) + w_{+1}(x) \psi(t,x+1).
$$
Here the weights $w_{\beta}(x) \in M_2(\C)$ are $2 \times 2$ complex
matrices constrained by the requirement that the global evolution 
matrix
$$
U := \pmatrix{ 
  \ddots   &             &             &             
           &             &             &             \cr
           & w_{-1}(x-1) &      0      & w_{+1}(x-1)  
           &             &             &             \cr
           &             &  w_{-1}(x)  &      0      
           &  w_{+1}(x)  &             &             \cr
           &             &             & w_{-1}(x+1) 
           &      0      & w_{+1}(x+1) &             \cr
           &             &             &
           &             &             &   \ddots    \cr
}                                                             \eqno(1)
$$
be unitary.

For periodic boundary conditions the top and bottom rows (of blocks) 
of $U$ wrap around and completely homogeneous solutions are possible.  
We showed in [1] that the most general parity invariant homogeneous 
solutions, up to unitary equivalence and an overall unobservable 
phase, form a one parameter family
$$
w_{-1}(x) = w_{-1} := \pmatrix{0 & i\sin\theta \cr
                               0 &  \cos\theta \cr
                              }
\qquad
w_{+1}(x) = w_{+1} := \pmatrix{ \cos\theta & 0 \cr
                               i\sin\theta & 0 \cr
                              },
$$
where $\theta$ scales to the mass in the Dirac equation limit.

Without periodic boundary conditions, the local rule must change at
the boundary.  We show in [11] that there is a one parameter family
of (Type II) boundary conditions which give the global evolution 
matrix the form:
$$
U := \pmatrix{   0   & \overline{w}_{+1} &        &        \cr
              w_{-1} &         0         & w_{+1} &        \cr
                     &      w_{-1}       &    0   &        \cr
                     &                   &        & \ddots \cr
             }                                                \eqno(2)
$$
at the left boundary.  Here 
$$
\overline{w}_{+1} := \pmatrix{     0      &  0  \cr
                              ie^{i\zeta} &  0  \cr
                             }
$$
for any $\zeta \in \R$ and the right boundary condition is 
characterized by $\overline{w}_{-1}$ obtained by parity 
transformation, although with an independent phase.  For simplicity we
set both phase angles to zero in the following discussion; this will 
not affect the relevant features of our results.

The global evolution matrix $U$ acts by left multiplication 
$\Psi(t+1) = U \Psi(t)$ on the wave function 
$\Psi(t) : L \times \{\pm1\} \longrightarrow \C$.  Multiplication of 
the wave function by an overall phase $e^{-i\alpha}$ leaves the 
quantum state invariant since the phase cancels in any observable 
$\langle\Psi|{\cal O}|\Psi\rangle$.  Requiring invariance of the 
evolution under the local gauge transformations $e^{-i\alpha(t,x)}$ 
corresponding to this global symmetry necessitates modification of 
$U$:  Setting $\psi'(t,x) := e^{-i\alpha(t,x)}\psi(t,x)$, which we 
write as 
$$
\eqalignno{
\Psi'(t) &:= D[e^{-i\alpha(t,x)}]\Psi(t),                     &(3a)\cr
\noalign{\smallskip}
\noalign{\hbox{we achieve gauge invariance with}} 
\noalign{\smallskip}
   U'(t) &:= D[e^{-i\alpha(t+1,x)}]UD[e^{i\alpha(t,x)}]       &(3b)\cr
}
$$
since
$$
\eqalign{
U'(t)\Psi'(t) &= D[e^{-i\alpha(t+1,x)}]UD[e^{i\alpha(t,x)}]
                 D[e^{-i\alpha(t,x)}]\Psi(t)                       \cr
              &= D[e^{-i\alpha(t+1,x)}]\Psi(t+1)                   \cr
              &= \Psi'(t+1).                                       \cr
}
$$

Expanding ($3b$) we find that the superdiagonal blocks of $U'(t)$ are
$$
\eqalignno{
U'(t)_{x-1,x} &= e^{-i\alpha(t+1,x-1)+i\alpha(t,x)}w_{+1},    &(4a)\cr
\noalign{\smallskip}
\noalign{\hbox{while the subdiagonal blocks are}}
\noalign{\smallskip}
U'(t)_{x,x-1} &= e^{-i\alpha(t+1,x)+i\alpha(t,x-1)}w_{-1}.    &(4b)\cr
}
$$
For a periodic lattice, $x$ runs over all the labels 
$\{0,\dots,|L|-1\}$ of the lattice sites and the indices are 
interpreted mod $|L|$.  For a lattice with boundaries, 
$x \in \{1,\ldots,|L|-1\}$ since there are no blocks in the evolution 
matrix containing nonzero amplitudes for transitions between $x = 0$ 
and $x = |L|-1$.  Multiplying and dividing each block in column $x$ 
(with these same conventions) of $U'(t)$ by $e^{i\alpha(t+1,x)}$, the 
expressions in (4) can be rewritten as
$$
\eqalignno{
U'(t)_{x-1,x} 
 &= e^{-i\Delta_t\alpha(t+1,x)  +i\Delta_x\alpha(t+1,x)}w_{+1}  
                                                              &(5a)\cr
\noalign{\hbox{and}}
U'(t)_{x,x-1}
 &= e^{-i\Delta_t\alpha(t+1,x-1)-i\Delta_x\alpha(t+1,x)}w_{-1},
                                                              &(5b)\cr
}
$$
where $\Delta$ is the difference operator so 
$\Delta_t\alpha(t+1,x) := \alpha(t+1,x) - \alpha(t,x)$ and similarly 
for $\Delta_x$.  In column $x$ both the super- and subdiagonal blocks
now contain the phase factor $e^{-i\Delta_t\alpha(t+1,x)}$.  Recall 
that when we included an inhomogeneous potential in a QLGA [12] it
also had the effect of multiplying the blocks in column $x$ by a phase
factor, $e^{-i\phi(x)}$.%
\sfootnote*{Column and row are interchangeable here, corresponding to
multiplying by the local potential phase factors before or after the 
advection part of each timestep, respectively.  The algorithm 
described in [12] multiplies before, while those in [9] and [13] do
so afterwards.  To obtain the expressions corresponding to those in 
(5), but with a common $\Delta_t\alpha$ phase in each row, we can 
multiply and divide each block in row $x$ of $U'(t)$ by 
$e^{i\alpha(t,x)}$.}
Thus the gauge transformation (3) transforms a potential $\phi(t,x)$ 
to
$$
\phi'(t,x) = \phi(t,x) + \Delta_t\alpha(t+1,x).               \eqno(6)
$$

The $\Delta_x\alpha$ phases in the gauge transformed evolution matrix
$U'(t)$ remind us that the electromagnetic potential has a second
component---the vector potential $A$.  The conjugate $\Delta_x\alpha$
phases of the antidiagonal blocks in (5) indicate that the vector 
potential enters the evolution matrix in the same way.  That is, in 
the presence of an electromagnetic potential 
$\bigl(\phi(t,x),A(t,x)\bigr)$, the super- and subdiagonal blocks in 
the evolution matrix are
$$
\eqalignno{
U^{(\phi,A)}(t)_{x-1,x} &= e^{-i\phi(t,x)+iA(t,x)}w_{+1}      &(7a)\cr
\noalign{\hbox{and}}
U^{(\phi,A)}(t)_{x,x-1} &= e^{-i\phi(t,x)-iA(t,x)}w_{-1},     &(7b)\cr
}
$$
respectively.  Notice that for a lattice with boundaries, 
$A(t,0) \equiv A(t,|L|)$ does not enter into these expressions for the
global evolution, although it does for a periodic lattice.  
$U^{(\phi,A)}$ describes the discrete version of the usual minimal 
coupling of a Dirac particle to an\break

\moveright\secondstart\vtop to 0pt{\hsize=\halfwidth
\null\vbox to\vsize{\vfill
$$
\epsfxsize=\halfwidth\epsfbox{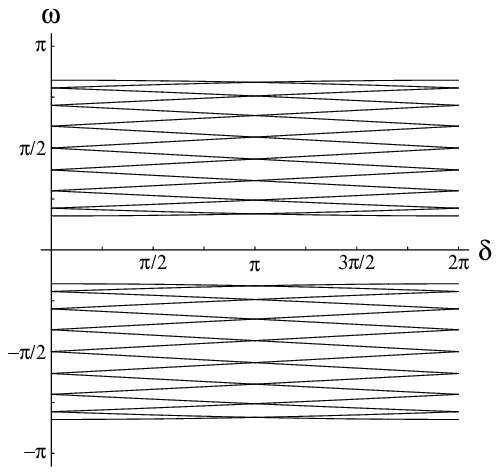}
$$
\vskip-5\baselineskip
\eightpoint{%
\noindent{\bf Figure 1}.  Spectral flow in a periodic QLGA with 
$|L| = 16$ and $\theta = \pi/6$.  The scalar potential has been taken 
to vanish and $\delta$, the gauge invariant degree of freedom in the 
vector potential runs from $0$ to $2\pi$.
\vskip 11pt
}}}
\vskip -2\baselineskip
\noindent  external (classical)
electromagnetic field [14].  The conjugate phases from the vector
potential are the consequence of the velocity dependence of this
coupling; the vector potential itself transforms to
$$
A'(t,x) = A(t,x) + \Delta_x\alpha(t+1,x)                      \eqno(8)
$$
under the gauge transformation (3), analogously to the transformation 
(6) of the scalar potential.

\parshape=18
0pt \hsize
0pt \hsize
0pt \hsize
0pt \hsize
0pt \hsize
0pt \hsize
0pt \hsize
0pt \hsize
0pt \hsize
0pt \hsize
0pt \hsize
0pt \hsize
0pt \hsize
0pt \hsize
0pt \hsize
0pt \hsize
0pt \hsize
0pt \halfwidth
The consequences of this gauge freedom in the vector potential differ
for periodic lattices and those with boundaries:  Given any vector
potential $A(t,x)$, consider the gauge transformation defined by
$$
\alpha(t+1,x) = -\sum_{y=0}^x A(t,y).                         \eqno(9)
$$
With this definition, $\Delta_x\alpha(t+1,x) = -A(t,x)$ for 
$1 \le x \le |L|-1$ so that using (9) in (8) gives a gauge transformed
$A'(t,x)$ which vanishes except at $x = 0$.  Since the global 
evolution on a lattice with boundaries does not depend on $A'(t,0)$,
(9) gauge transforms the vector potential away completely---there is
exactly one gauge equivalence class of vector potentials.  On a 
periodic lattice, however, the global evolution depends on $A'(t,0)$
(mod $2\pi$), so there is a one (periodic) parameter family of gauge
equivalence classes.  Alternatively, the gauge transformation (9)
demonstrates that the only gauge invariant quantity which can be
constructed from the vector potential is (a function of)
$$
\delta := \exp i\sum_{x=0}^{|L|-1} A(t,x),                   \eqno(10)
$$
upon which the evolution depends only for a periodic lattice, in which
case (10) is the discrete analogue of a Wilson loop variable/holonomy,
\ie, a winding number.

\parshape=15
0pt \halfwidth
0pt \halfwidth
0pt \halfwidth
0pt \halfwidth
0pt \halfwidth
0pt \halfwidth
0pt \halfwidth
0pt \halfwidth
0pt \halfwidth
0pt \halfwidth
0pt \halfwidth
0pt \halfwidth
0pt \halfwidth
0pt \halfwidth
0pt \hsize
Such a gauge invariant quantity manifests itself in the spectrum of 
the evolution operator:  As the phase $\delta$ varies, so does the 
spectrum of $U^{(\phi,A)}$---if the lattice is periodic, but 
{\sl not\/} if it has boundaries.  Figure~1 shows an example of this 
spectral flow [10] for a periodic lattice with $N = 16$ and 
$\theta = \pi/6$.  Here $\phi(t,x)$ has been set to 0 by a gauge 
transformation, and $\delta$ varies from $0$ to $2\pi$.  As $\delta$ 
increases, the frequencies (energies) of positive frequency right/left 
moving plane waves, \ie, the eigenvectors of 
$U^{(\phi,A)}$, increase/decrease, respectively.  When the `mass'
$\theta = 0$, half the eigenvalues shift up by 1 and half shift down
by 1 as $\delta$ varies from 0 to $2\pi$; the spectral flow (the  
number of eigenvalues crossing each frequency level) is 2.

To observe such a spectral flow, imagine preparing a quantum particle 
on a periodic lattice in a positive frequency, right moving eigenstate
of $U^{(0,0)}$, but then applying an external vector potential so that
the evolution is by $U^{(0,A)}$ instead, where $A(t,x) \equiv A$.  A
massless particle will remain in the original eigenstate, but a 
frequency (energy) measurement at any subsequent timestep will return
the original value plus $A = \delta/|L|$.  For a massive (\ie, 
$\theta \not= 0$) particle, the initial state is a superposition of
the positive and negative frequency eigenstates of $U^{(0,A)}$ with 
the original wave number, but for small masses the positive frequency
still dominates so again a frequency (energy) measurement will return
the original value plus an amount on the order of $A$.

Our goal, however, is to distinguish between a periodic lattice and 
one with boundaries.  Since the single particle eigenstates differ for 
these two situations even if $|L|$ is the same [13,11], without
knowing the spatial topology {\it a priori\/} we cannot prepare a 
quantum particle in an eigenstate.  But we can prepare a positive 
frequency right moving wave packet with bounded support---identically 
on {\sl either\/} lattice.  Subsequent frequency measurements will 
return a distribution of frequencies concentrated around the 
expectation value.  This distribution is the same for either lattice, 
if the evolution is by $U^{(0,0)}$.  But if the evolution is by 
$U^{(0,A)}$ with $A > 0$, say, the observed frequency distribution 
will be shifted to larger values on the periodic lattice.

Since some finite, fixed, number of measurements suffices to 
distinguish these distributions with probability $1 - \epsilon$ for a
given initial wave packet and external vector potential, the 
computation takes only {\sl constant\/} time, independent of the size 
of the lattice $|L|$.  In fact, for larger lattices the initial wave 
packet can be broader, \ie, more concentrated around its expected 
energy, and hence {\sl fewer\/} measurements are needed to distinguish 
the original distribution from the one shifted by the vector potential 
in case the lattice is periodic.  In contrast, suppose we try to 
distinguish a periodic lattice from one with boundaries using a 
{\sl deterministic\/} LGA.  Since no superpositions of states are 
possible, all that we can do is to start a single particle off in one
direction and see if it ever changes velocity by reflecting from a 
boundary.  To do so the particle must travel for $O(|L|)$ timesteps,
on the average, so the QLGA algorithm provides a even greater 
improvement over the classical algorithm than does Grover's algorithm
for searching [4].

In conclusion, we remark that our algorithm exploits the Aharonov-Bohm
effect [15] which is more usually discussed in two dimensions---we are
imagining creating the vector potential by applying a magnetic field 
which threads a (possibly incomplete, in the non-periodic case) ring 
of lattice sites.  It would be interesting, and potentially useful for 
pattern recognition [16], to formulate this algorithm for a two 
dimensional QLGA.  Preparation of a localized wave packet, and 
measurement of its energy, each in constant time, is then plausible on 
physical grounds, provided that the lattice lies within some fixed 
area.  For the more realistic situation of a fixed density of lattice 
sites (and hence an increasing spatial area with increasing lattice 
size) additional analysis is required.  As the example of unitary 
transformations and measurements on the Rydberg states of an atom
illustrates, careful attention to the details of the physical 
implementation of a quantum algorithm is required to correctly 
quantify its computational complexity [17].  Although the evolution of
the QLGA can be simulated in the standard poly-local model for quantum
computing [18], issues of the time required for state preparation and 
measurement, and the adiabaticity required when turning on the 
magnetic field, all must be considered before we can claim to 
understand how this computation scales physically.

\bigskip
\noindent{\bf Acknowledgements}

\noindent I thank A. P. Balachandran for encouraging me to think about
possible QLGA algorithms, Peter Doyle, Mike Freedman and Brosl 
Hasslacher for useful discussions, and Sun Microsystems for providing 
computational support.  This work was supported in part by the 
National Security Agency (NSA) and Advanced Research and Development 
Activity (ARDA) under Army Research Office (ARO) contract number 
DAAG55-98-1-0376.

\bigskip
\global\setbox1=\hbox{[00]\enspace}
\parindent=\wd1

\noindent{\bf References}
\bigskip

\parskip=0pt
\item{[1]}
\dajm,
``From quantum cellular automata to quantum lattice gases'',
\JSP\ {\bf 85} (1996) 551--574.

\item{[2]}
B. Meurer, D. Heitmann and K. Ploog,
``Single-electron charging of quantum-dot atoms'',
\PRL\ {\bf 68} (1992) 1371--1374;\hfb
G. Springholz, M. Pinczolits, P. Mayer, V. Holy, G. Bauer, H. H. King
and L. Salamanca-Riba,
``Tuning of vertical and lateral correlations in self-organized
  PbSe/ Pb$_{1-x}$Eu$_x$Te quantum dot superlattices'',
\PRL\ {\bf 84} (2000) 4669--4672;\hfb
and references therein.

\item{[3]}
P. W. Shor,
``Algorithms for quantum computation:  discrete logarithms and 
  factoring'',
in S. Goldwasser, ed.,
{\sl Proceedings of the 35th Symposium on Foundations of Computer 
Science}, Santa Fe, NM, 20--22 November 1994
(Los Alamitos, CA:  IEEE Computer Society Press 1994) 124--134.

\item{[4]}
L. K. Grover,
``A fast quantum mechanical algorithm for database search'',
in
{\sl Proceedings of the 28th Annual ACM Symposium on the Theory 
of Computing}, Philadelphia, PA, 22--24 May 1996
(New York:  ACM 1996) 212--219.

\item{[5]}
See, \eg,
D. Beckman, A. N. Chari, S. Devabhaktuni, J. Preskill,
``Efficient networks for quantum factoring'',
\PRA\ {\bf 54} (1996) 1034--1063.

\item{[6]}
N. Margolus,
``Physics-like models of computation'',
\PD\ {\bf 10} (1984) 81--95;\hfb
M. Biafore,
``Cellular automata for nanometer-scale computation'',
\PD\ {\bf 70}\break
(1994) 415--433.

\item{[7]}
\brosl,
``Discrete fluids'',
\LAS\ {\bf 15} (1988) 175--200, 211--217.

\item{[8]}
\bogtay,
``Simulating quantum mechanics on a quantum computer'',
\PD\ {\bf 120} (1998) 30--42.

\item{[9]}
\bogtay,
``A quantum lattice-gas model for the many-particle Schr\"odinger
  equation in $d$ dimensions'',
\PRE\ {\bf 8} (1997) 705--716.

\item{[10]}
M. F. Atiyah, V. K. Patodi and I. M. Singer,
``Spectral asymmetry and Riemannian geometry.  III'',
\MPCPS\ {\bf 79} (1976) 71--99.

\item{[11]}
\dajm,
``Quantum mechanics of lattice gas automata:  
  Boundary conditions and other inhomogeneities'',
\JPA\ {\bf 31} (1998) 2321--2340.

\item{[12]}
\dajm,
``Quantum lattice gases and their invariants'',
\IJMPC\ {\bf 8} (1997) 717--735.

\item{[13]}
\dajm,
``Quantum mechanics of lattice gas automata:  
  One-particle plane waves and potentials'',
\PRE\ {\bf 55} (1997) 5261--5269.

\item{[14]}
P. A. M. Dirac,
``The quantum theory of the electron'',
\PRSLA\ {\bf 117} (1928) 610--624.

\item{[15]}
Y. Aharonov and D. Bohm,
``Significance of electromagnetic potentials in the quantum 
  theory'',
\PR\ {\bf 115} (1959) 485--491.

\item{[16]}
T. Y. Kong and A. Rosenfeld, eds.,
{\sl Topological Algorithms for Digital Image Processing\/}
(New York:  Elsevier 1996);\hfb
and references therein.

\item{[17]}
D. A. Meyer, P. G. Kwiat, R. J. Hughes, P. H. Bucksbaum, J. Ahn and 
T. C. Weinacht,
``Does Rydberg state manipulation equal quantum computation?'',
\Sc\ {\bf 289} (1 sep 2000) 1431a.

\item{[18]}
M. H. Freedman,
``Poly-locality in quantum computing'',
{\tt quant-ph/0001077}.

\bye